\documentclass[preprint,amsmath,amssymb]{revtex4}

\usepackage{latexsym}
\usepackage[dvips]{graphicx}
\usepackage{amsmath,amssymb}

\usepackage[dvips]{graphicx}
\usepackage{amsmath,amssymb}

\newcommand{\1}{\'{\i}}

\newcommand{\beq}{\begin{equation}}
\newcommand{\eeq}{\end{equation}}
\newcommand{\ba}{\begin{array}}
\newcommand{\ea}{\end{array}}
\newcommand{\beqa}{\begin{eqnarray}}
\newcommand{\eeqa}{\end{eqnarray}}

\begin{document}

\title{Chromomagnetic Dipole Moment of the Top Quark Revisited}
\author{R. Mart\'{\i}nez(1), M. A. P\'erez(2) and N. Poveda T.(1,3)}
\altaffiliation{remartinezm@unal.edu.co, mperez@fis.cinvestav.mx, npoveda@tunja.uptc.edu.co}
\affiliation{(1) Departamento de F\'\i sica, Universidad Nacional de Colombia \\
Bogot\'a, Colombia\\
(2) Departamento de F\'\i sica, Cinvestav - IPN,\\
Av. IPN No. 2508, M\'exico\\
(3) Departamento de F\'\i sica, Universidad Pedag\'ogica y Tecnol\'ogica de Colombia\\
Tunja, Colombia}

\date{}

\begin{abstract}
We study the complete one-loop contributions to the chromagnetic
dipole moment $\Delta\kappa$ of the top quark in the Standard Model,
two Higgs doublet models, topcolor assited technicolor models
(TC2), 331 models and extended models with a single extra
dimension. We find that the SM predicts $\Delta\kappa = - 0.056$ and
 the predictions of the other models are also consitent with
the constraints imposed on $\Delta\kappa$ by low-energy precision
measurements.
\end{abstract}
\maketitle

\section{Introduction}

The fact that the top quark mass $\sim 172 GeV$ \cite{uno} is of
the same order of magnitude then the electroweak symmetry breaking
(EWSB) scale $\nu = (\sqrt 2 G_F)^{-1/2} = 246$ GeV suggests that
the top quark may be more sensitive to new physics effects than
the remaining ligthter fermions. With the advent of the CERN Large
Hadron Collider (LHC), 8 millions top-quark pairs will be produced
per year  with an integrated luminosity of $10 fb^{-1}$
\cite{dos}. This number will increase by one oder of magnitude
with the high luminosity option $(100 fb^{-1})$. Therefore, the
properties of the top quark will be examined with significant
precision at the LHC. In particular, the interest in non-standard
$t t g$ couplings arised some time ago when it was realized that
the presence of non-Standard-Model couplings could lead to
significant modifications in the total and differential cross
sections of top-pair production at hadron colliders
\cite{tres,cuatro}. If the source of this new physics is at the
TeV scale, it has been pointed out \cite{tres} that the leading
effect may be parametrized by a chromomagnetic dipole moment
(CMDM) $\Delta\kappa$ of the top quark since this is the lowest
dimension CP-conserving operator arising from an effective
Lagrangian contributing to the gluon-top-quark coupling,
\begin{equation}
{\cal L}_5 = (\Delta\kappa/2)(g_s/4m_t)\bar
u(t)\sigma_{\mu\nu}F^{\mu\nu, \alpha} T^\alpha u(t)
\end{equation}where $g_s$ and $T^\alpha$ are the $SU(3)_c$
coupling and generators, respectively. $F^{\mu\nu,
\alpha}=\partial^\mu A^{\nu, \alpha} - \partial^\nu A^{\mu,
\alpha} -g_s f^{\alpha\beta\gamma} A^{\mu, \beta} A^{\nu, \gamma}$
is the gluonic antisymmetric tensor.

The effects due to $\Delta\kappa \neq 0$ have been examined in
flavor physics as well as in topquark cross section measurements
\cite{tres,cuatro,cinco}. In the latter case, the parton level
differential cross sections of $gg \to \bar tt$ and $\bar q q \to
\bar t t$ (the dominant channel at Fermilab Tevatron energies)
were calculated \cite{tres,cuatro,seis}. The combined effects of
the chromomagnetic and the chromoelectric dipole moment of the top
quark on the reaction $p\bar p \to t\bar t X$ were investigated
in Ref. \cite{cinco}. Moreover, previous analysis has revealed
the the differential cross section is sensitive to the sign of
the anomalous chromomagnetic dipole moment on account of the
interference with the SM coupling. This can lead to a significant
suppression or enhancement in the production rate \cite{tres}.

Cross section measurements at the Tevatron are expected to
constrain the CMDM of the top quark to $|\Delta\kappa| \leq 0.15
- 0.20$ \cite{siete}. Since the influence of $\Delta\kappa$ grows
rapidly with the increasing center of mass energy, this bound
will be improved by one order of magnitude at the LHC with a
luminosity of $100 fb^{-1}$ \cite{ocho}. On the other hand, it has
been pointed out \cite{nueve} that the CLEO data on $b \to
s\gamma$ gives already a constraint as strong as that expected at
the LHC, $-0.03 \leq \Delta\kappa \leq 0.01$. Low-energy precision
measurements have produced similar constraints for the
non-standard top-quark couplings $tbW, ttZ, tcV, tcH$, with $V =
\gamma, g, Z$ \cite{diez, once}. It is interesting to notice that
the CP violating chromoelectric dipole moment of the top quark,
which is much further suppressed than the CMDM, has not been
constrained yet by low-energy precision measurements. However,
the CP-odd observables induced by a chromoelectric dipole moment
for the $t\bar t$ system have been studied in $p-p$ and $p-\bar
p$ collisions \cite{doce}.

In the Standard Model (SM) \cite{trece}, $\Delta\kappa$ arises at
the one-loop level and it is of order $10^{-2}$ for a light Higgs
boson mass \cite{nueve}. A large value for $\Delta\kappa$ arises
naturally in dynamical electroweak-symmetric breaking models such
as technicolor or topcolor \cite{tres}. In two Higgs doublet
models (THDM) and for QCD-SUSY corrections, previous studies have
found that $\Delta\kappa$ could be as large as $10^{-1}$
\cite{nueve}. The search for $\Delta\kappa$ effects induced in the
LHC/ILC accelerators constitute then a window to look for physics
beyond the SM.

Motivated by the fact that there is no detailed study in the
literature of the top-quark CMDM $\Delta\kappa$,  in the present
paper we make a critical reanalysis of the one-loop contributions
to $\Delta\kappa$ in the SM, the two-Higgs doublet model
(THDM-II), topcolor assited technicolor (TC2) models, as well as
in the so called 331 models and in the framework of a universal
extra dimension with the SM fields propagating in the bulk. We
have found that the predictions of all these models are
consistent with the constraints obtained from low-energy
precision measurements \cite{nueve}.

The paper is organized as follows. In Section II we present the
general framework required in each model in order to compute the
respective one-loop contributions to the top-quark CMDM. In
this paper we put together the results obtained for the
top-quark CMDM in each one of the models addressed in this paper.
The concluding remarks are included in section III and in the
Appendix we present the analytical expressions obtained in our
calculation for the one-loop Feynman diagrams contributing to the
CMDM.

\section{Framework}
In this section, we present the basic elements of the models in
which we have computed the top-quark CMDM.
\subsection{Standard Model}
\begin{figure}
  \includegraphics[width=10cm]{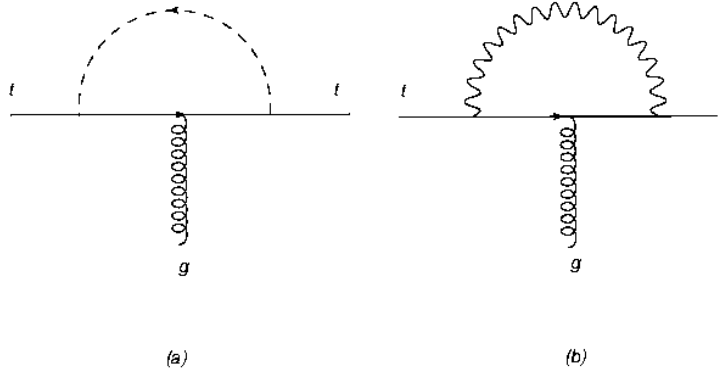}\\
  \caption{Feynman diagrams that contribute to the CMDM}\label{digfey1}
\end{figure}

\begin{figure}
 \includegraphics[width=10cm]{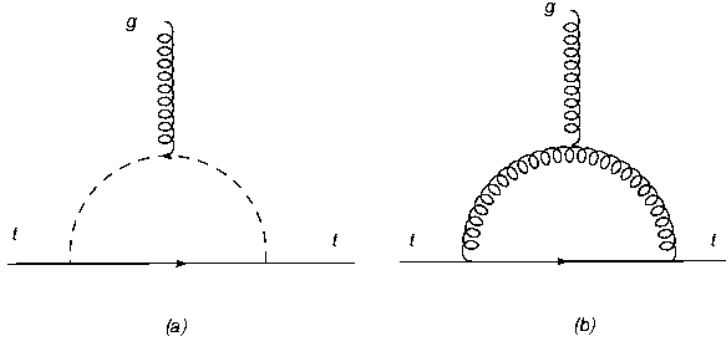}\\
  \caption{Feynman diagrams for the QCD contributions to the CMDM}\label{digfey2}
\end{figure}

In the SM \cite{diez} the top-quark CMDM arises from loops
containing the electroweak bosons $W^\pm, Z, h^0$ with their
respective would-be Goldstone bosons $G^\pm_W$ and $G_Z$, or just
gluons \cite{nueve}. Figures \ref{digfey1} and \ref{digfey2} show
the generic diagrams for the electroweak contributions to the
CMDM. In Table I we include the respective Feynman rules used to
compute these contributions. The diagrams shown in Figs. 1b and
2b induce the QCD contribution to the CMDM of the top quark.
Notice that the contribution arising from diagram $2b$ was not
considered in our previous work on $\triangle\kappa$ \cite{nueve}.

\begin{table}
\begin{center}
\begin{tabular}{||l||c|c|c|c||}
  \hline\hline
              & $A$ & $B$ & $A^{'}$  & $B^{'}$  \\
\hline\hline
  $tth^0$ & $-i\frac{gmt}{2M_W}$ & $-i\frac{gmt}{2M_W}$ & $-i\frac{gmt}{2M_W}$ & $-i\frac{gmt}{2M_W}$ \\
  \hline
  $ttG_Z$ & $ \frac{gmt}{2M_W}$ & $ -\frac{gmt}{2M_W} $ & $ \frac{gmt}{2M_W} $ &  $ -\frac{gmt}{2M_W} $  \\
  \hline
  $ttG_W$ & 0 & $ i\frac{gmt}{2M_W}$ & $i \frac{gmt}{2M_W} $ & 0 \\
  \hline
  $tbW_\mu$ & $-i\frac{g}{\sqrt{2}} $ & 0 & $-i\frac{g}{\sqrt{2}}$  &  0  \\
 \hline
  $ttZ_\mu$ & $ -i\frac{g}{c_W}(\frac{1}{2}-\frac{2}{3}s_W^2) $ &  $i\frac{g}{c_W}\frac{2}{3}s_W^2 $ & $-i\frac{g}{c_W}(\frac{1}{2}-\frac{2}{3}s_W^2) $ &  $ i\frac{g}{c_W}\frac{2}{3}s_W^2$  \\
  \hline\hline
\end{tabular}
\end{center}
\caption{Feynman rules for the SM contributions to the
CMDMT}\label{sm}\end{table}

The contributions obtained for each SM QCD and electroweak one-loop
diagrams for the CMDM are given by

\begin{eqnarray}
\Delta\kappa_{(\mu)}^{QCD}&=&-6.4\times 10^{-2}  \nonumber \\
\Delta\kappa_{(s)}^{h^0}&=& 8.7 \times 10^{-3}\;\;\;\; ,\;\;\; m_h=120 GeV\nonumber \\
\Delta\kappa_{(s)}^{G_Z}&=& -5.1 \times 10^{-3} \nonumber \\
\Delta\kappa_{(s)}^{G_W}&=& -5.2 \times 10^{-3} \nonumber \\
\Delta\kappa_{(\mu)}^{W}&=& 9.9 \times 10^{-3} \nonumber \\
\Delta\kappa_{(\mu)}^{Z}&=& -7.5 \times 10^{-4}  \nonumber \\
\end{eqnarray}
where the subindices $(\mu)$ or $(s)$ mean that the internal boson
line in the one loop diagram  corresponds to gauge or scalar bosons,
respectively. Taking the sum of only the electroweak contributions,
we get
\begin{equation}
\Delta\kappa^{EW}= 7.5 \times 10^{-3}
\end{equation}which is about $12\%$ of the QCD contribution but with opposite
sign.  The SM prediction for the top-quark CMDM is thus given by
\begin{eqnarray}
\Delta\kappa^{SM} = - 5.6 \times 10^{-2}
\end{eqnarray}

\subsection{THDM-II}
The Yukawa couplings needed to compute the top-quark CMDM in the
two Higgs doublet model (THDM-II) are given by the Lagrangian
\beq -{\cal L}_Y= h_t \overline{Q}_L \widetilde{\phi}_1 t_R + h_b
\overline{Q}_L \phi_2 b_R + h.c. \eeq where we have used a
discrete symmetry $\phi_1 \to  \phi_1 \to \phi_2 \to - \phi_2$ to
avoid flavor-changing neutral couplings for the quarks at the
tree level \cite{catorce}.  $Q_L$ corresponds to the quark doublet
for the third family. After the spontaneous symmetry breaking,
with $\nu_1$  and $\nu_2$ the respective vacuum expectation
values for the two Higgs doublets and $\tan \beta = \nu_1/\nu_2$,
there are two physical charged scalar bosons $H^\pm$ and three
physical neutral scalar bosons $h, H$ and $A$, and the respective
would-be Goldstone bosons $G^\pm_W$ and $G_Z$. The Feynman rules
required to compute the top-quark CMDM in this model are depicted
in Table II. We have neglected the $H^\pm tb$ coupling which is
proportional to the bottom quark mass. The $\cos\theta$ is the
mixing angle of the two CP-even scalar fields, $h, H$.   
  
The one-loop effects induced by the THDM on the CMDM were also
calculated in Ref.[9]. However, the results obtained in this
case were used only to get constraints on the THDM parameters
involved in the calculation by requiring in turn that these
contributions also agreed with the low-energy constraints on
the CMDM \cite{nueve}.

\begin{table}
\begin{center}
\begin{tabular}{||l||c|c|c|c||}
  \hline\hline
              & $A$ & $B$ & $A^{'}$  & $B^{'}$  \\
\hline\hline
  $tth$ & $-i\frac{gm_tc_{\theta}}{2M_Ws_{\beta}}$ & $-i\frac{gm_tc_{\theta}}{2M_Ws_{\beta}}$ & 
$-i\frac{gm_tc_{\theta}}{2M_Ws_{\beta}}$ &  $-i\frac{gm_tc_{\theta}}{2M_Ws_{\beta}}$ \\
  \hline
  $ttA^0$ & $ \frac{gm_t}{2M_Wt_{\beta}}$ & $ -\frac{gm_t}{2M_Wt_{\beta}} $ & $ \frac{gm_t}{2M_Wt_{\beta}} $ &  $ -\frac{gm_t}{2M_Wt_{\beta}} $  \\
  \hline
  $tbH$ & $0 $ & $i\frac{gm_t}{\sqrt{2}M_Wt_{\beta}}$ & $i\frac{gm_t}{\sqrt{2}M_Wt_{\beta}}$  &  0  \\
  \hline\hline
\end{tabular}
\end{center}
\caption{Feynman rules for the THDM contributions to the
CMDMT}\label{THDM}
\end{table}

In the THDM-II, the charged Higgs boson $H^+$ and the neutral Higgs
bosons $h^0, H^0, A^0$ give the following contributions to the CMDM,
\begin{eqnarray}
\Delta\kappa^{H^+}_{(s)}&=& 3.7 \times 10^{-3} ,\;\; 1.4 \times
10^{-3},\;\; 6.2 \times 10^{-4}\;\; , \nonumber \\
\Delta\kappa^{A^0}_{(s)}&=&-3.6 \times 10^{-3} ,\;\; -2.8 \times
10^{-3},\;\; -1.8 \times 10^{-3}, \nonumber \\
\Delta\kappa^{h^0}_{(s)}&=& 8.6 \times 10^{-3}, \nonumber \\
\Delta\kappa^{H^0}_{(s)}&=& 4.3 \times 10^{-3},\;\;
\end{eqnarray}
where we have taken  $\tan \beta = 1$, $\sin\theta = 1/\sqrt{2}$ and
$m_{H^+}/m_{A^0} = 250, 350, 500$ GeV. The masses for the CP-even
scalar Higgs $h^0, H^0$ are $120, 300$ GeV, respectively. We observe
that these contributions become smaller with heavier scalars, which
agrees with the expectations given by the decoupling theorem. Taking
into account the contribution of the $H^+, H^0, A^0$  fields, we get
for the scalar contributions of the THDM-II to the CMDM the result
\begin{equation}
\Delta\kappa^{THDM}_{(s)}= 4.4 \times 10^{-3}, \;\; 3.0 \times
10^{-3}, \;\; 3.1 \times 10^{-3}.
\end{equation}
Then, the THDM contribution to the CMDM is one order of magnitude
smaller than the SM value.

When  $\tan\beta=10$ the predicted value for $\Delta\kappa$ due
to the CP-even neutral scalar is of the same order of magnitude than for
$\tan\beta=1$,
but the CP-odd scalar contribution is two orders of magnitude lower,
and the charged scalar Higgs is  one order of magnitude lower than
the value given by $\tan\beta=1$. In this case the most important
contribution is coming from the $H^0$ scalar field, and the
$\Delta\kappa^{THDM}$ practically does not change for this value of
$\tan\beta$. However, for a bigger value of this parameter
$\tan\beta=100$, the term proportional to the bottom quark mass in
the $tbH^+$ vertex is more important than the coupling proportional
to the top quark mass. The contributions coming from the CP-even Higgs
fields are unchanged and  the CP-even neutral scalar is very much
supppresed, but the charged Higgs is incresed in one order of
magnitude, i.e., for $m_{H^+}=250$ GeV the CMDM is of the order of
$3\times 10^{-2}$ which is as big as the SM value but with opposite
sign.

\subsection{The 331 model}

In the so-called 331 models, which are based on the $SU(3)_c
\otimes SU(3)_L \otimes U(1)$ gauge group \cite{quince,
dieciseis}, the cancellation of anomalies requires to have three
fermion families. The number of families in these models is
regulated by the values of the parameter $\beta$ given in the
definition of the respective electromagnetic charge
\cite{quince}.  We will consider the 331 model with $\beta = -
\sqrt 3$ which has a new exotic quark $J_3$ with electric charge
$\pm 4/3$ and the bilepton gauge bosons $X^{\pm\pm}$ with  masses
$M_X > 850$ GeV \cite{quince}. The third family of quarks is given
by

\begin{eqnarray}
 Q_{3L} &=& \left( \ba{c} t \\ b\\ J_3 \ea \right)_L \sim (3,3,2/3) \ ;
 \nonumber \\
 t_{R} &\sim &(3,1,2/3) \ ; \ b_{R} \sim (3,1,-1/3) \ ; \
 J_{3R} \sim (3,1,4/3) \;\; ; \;\;
\end{eqnarray}
and the electric charge is defined by
\begin{equation}
Q=T_{3L}-\sqrt{3} T_{8L}+ X,
\end{equation}
where  $T_{3L}, T_{8L}$, and $X$ are the respective generators of
the groups $SU(3)_L$ and  $U(1)_X$.

The Higgs sector necessary to generate the fermionic masses is
given by three Higgs triplets, which after EWSB reduce to
 \beqa
\rho &=& \left( \ba{c} G_W^+ \\
\frac{i G_Z+V}{\sqrt{2}} \\ 0 \ea \right)  \sim (1,3,1)\;\ ; \;\
 \eta = \left( \ba{c} \frac{- i G_Z+V}{\sqrt{2}}
  \\  -G_W^-\\ 0 \ea \right) \sim (1,3,0)\; ; \;\nonumber \\
\chi &=& \left( \ba{c} G^-_Y \\ G^{--}_X \\
 \frac{w+i G_{Z'}}{\sqrt{2}} \ea \right)  \sim (1,3,-1),
\eeqa where $V$ and $\omega$ are the respective vacuum expectation
values and are chosen to obey the relation $\omega >\!> V$. The
scalar fields $G^\pm_W, G_Z, G^\pm_Y$ and $G_X^{\pm\pm}$
correspond to the would-be Goldstone bosons for the gauge fields
$W^\pm, Z, Y^\pm$ and $X^{\pm\pm}$, respectively.

The covariant derivative may be written in terms of the mass
eigenstates in the following way:
\beqa D_\mu &\sim& \partial_\mu + i \frac{g}{\sqrt{1-3 t_W^2}}
\left( \sqrt{3} t_W^2 (Q-T_3) +
T_8\right) Z'_\mu  \nonumber\\
&+& Y_\mu^- (T_4 - i T_5) + X_\mu^{--} (T_6 - i T_7) + h.c. \eeqa

Finally, the 331 Yukawa Lagrangian is given by
\beq -{\cal{L}}_Y =
 \lambda_3 \,\bar{Q}_{3L}\, J_{3R}\, \chi + \lambda_{ij}\,
\bar{Q}_{iL} \,J_{jR} \,\chi^*  + h.c.
  \eeq
where the coupling constants $\lambda_{33} $ and $\lambda_3$  are
given by
\beq \lambda_{33} \sim  \frac{g}{\sqrt{2}} \frac{m_b}{M_W}
\;\;\;\;,\;\;\;\; \lambda_3 \sim \sqrt{2} g \frac{m_{J_3}}{M_X}.
\eeq The relevant Feynman rules necessary to compute the
top-quark CMDM are given in Table III.

In the figure 3 we show the main contribution of this model to the
CMDM as a function of the $X$ gauge boson mass for three different
values of the exotic quark $J_3$ mass, $1, 2, 3$ TeV. Taking $3$ TeV
and $4$ TeV masses for $J_3$ and $X$, respectively  the CMDM is of
the order of $10^{-5}$ which is very much suppresed with respect to
the SM contribution.

\begin{figure}
  \includegraphics[width=10cm]{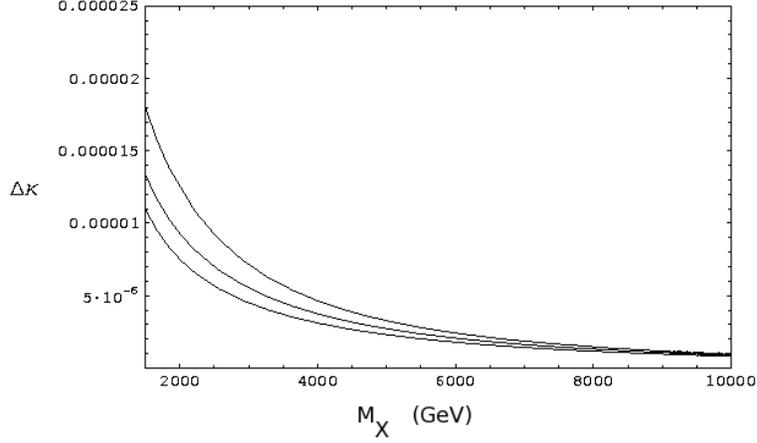}\\
  \caption{CMDM as a function of the bilepton gauge boson mass for three values of the $J_3$: 1, 2, 3 TeV.}\label{331}
\end{figure}

\begin{table}
\begin{center}
\begin{tabular}{||l||c|c|c|c||}
  \hline\hline
              & $A$ & $B$ & $A^{'}$  & $B^{'}$  \\
\hline\hline
  $tJ_3G_Y$ & $-i\frac{gm_J}{\sqrt{2}M_X}$ & 0 & 0 &  $-i\frac{gm_J}{\sqrt{2}M_X}$  \\
  \hline
  $tJ_3Y_\mu$ & $-i\frac{g}{\sqrt{2}} $ & 0 & $-i\frac{g}{\sqrt{2}}$  &  0  \\
 \hline
  $ttZ'_\mu$ & $ -i\frac{g(1+t_W^2)}{2\sqrt3\sqrt{1-3t_W^2}} $ & $ -i\frac{2g}{\sqrt3\sqrt{1-3t_W^2}} $ &
  $ -i\frac{g(1+t_W^2)}{2\sqrt3\sqrt{1-3t_W^2}} $ &  $ -i\frac{2g}{\sqrt3\sqrt{1-3t_W^2}} $ \\
  \hline
  $tJ_3X$ & $i \frac{g}{\sqrt{2}}$ & 0 & $i \frac{g}{\sqrt{2}}$ & 0 \\
  \hline\hline
  \end{tabular}

\end{center}
\caption{Feynman rules for the 331 model contributions to the
CMDM}\label{331}\end{table}

\subsection{Topcolor Assisted Technicolor}

Light particles of the SM can be regarded as spectators of the
electroweak symmetry breaking and the massive top quark suggests
that it is playing an important role in the dynamics. This implies
the possibility of a new interaction which drives the EWSB and the
big top mass in order to distinguish the top quark from the other
fermions. This interaction can generate desviations of the top quark
properties from the SM predictions.

In the topcolor scenario \cite{diecisiete,dieciocho,diecinueve,
veinte}, the EWSB mechanism arises from a new, strongly coupled
gauge interaction at TeV energy scales. In the TC2 model
\cite{diecisiete}, the topcolor interaction generates the top
quark condensation that gives rise to the main part of the
top-quark mass $(1-\varepsilon)m_t$, with the model dependent
parameter $\varepsilon$ fixed in the range $0.03 < \varepsilon <
0.1$ \cite{diecinueve}. This model predicts three heavy top-pions
$(\pi^0_t, \pi^\pm_t)$ and one top-Higgs boson $h^0_t$ with large
Yukawa couplings to the third generation of fermions. The
respective Yukawa couplings are obtained from the Lagrangian

\begin{equation}
{\cal L}=\mid D_\mu\Phi \mid^2-Y_t
\frac{\sqrt{v_w^2-F_t^2}}{v_w}\overline{\Psi}_L\Phi t_R- Y_t
\frac{\sqrt{v_w^2-F_t^2}}{v_w}\overline{t}_R\Phi \Psi_L -m_t
\overline{t}t,
\end{equation}
with $Y_t = (1-\varepsilon)m_t/F_t$, $\nu_\omega =\nu/\sqrt 2
\approx 174 GeV$, and the scalar field $\Phi$ is given by

\begin{equation}
\Phi=\left(\begin{array}{c} \frac{1}{\sqrt2}(h^0_t+i\pi_t^0) \\
\pi_t^+
\end{array}\right).
\end{equation}

For the purpose of the present paper, we will take the following
values for the topcolor parameters: $\varepsilon = 0.1$, $\nu_\omega
= 174$ GeV and $F_t = 50$ GeV. The Feynman rules for this model are
depicted in Table IV.

\begin{table}
\begin{center}
\begin{tabular}{||l||c|c|c|c||}
  \hline\hline
              & $A$ & $B$ & $A^{'}$  & $B^{'}$  \\
\hline\hline
  $tth_t^0$ & $-i\frac{(1-\epsilon)m_t}{\sqrt2v_w}F$ & $-i\frac{(1-\epsilon)m_t}{\sqrt{2}v_w}F$ &
  $-i\frac{(1-\epsilon)m_t}{\sqrt{2}v_w}F$ & $-i\frac{(1-\epsilon)m_t}{\sqrt{2}v_w}F$ \\
  \hline
  $tt\pi_t^0$ & $-\frac{(1-\epsilon)m_t}{\sqrt2v_w}F$ & $\frac{(1-\epsilon)m_t}{\sqrt2v_w}F$ &
  $-\frac{(1-\epsilon)m_t}{\sqrt2v_w}F$ &  $\frac{(1-\epsilon)m_t}{\sqrt2v_w}F$ \\
  \hline
  $tb\pi_t^+$ & $0$ & $-i\frac{(1-\epsilon)m_t}{v_w}F$ &  $-i\frac{(1-\epsilon)m_t}{v_w}F$
    &  0  \\
  \hline\hline
\end{tabular}

\end{center}

\caption{Feynman rules for the TC2 model contributions to the CMDM,
with $F=\sqrt{\frac{v_w^2}{F_t^2}-1}$}\label{TC2}\end{table}

The masses of the top-Higgs scalars $\pi^0_t$ and $\pi^+_t$  are
almost degenerate since they differ only by small electroweak
corrections \cite{veinte}. We will take for the top-Higgs mass a
value bigger than twice the value of the mass of the top-quark
\cite{veinte}.

\begin{figure}
  \includegraphics[width=10cm]{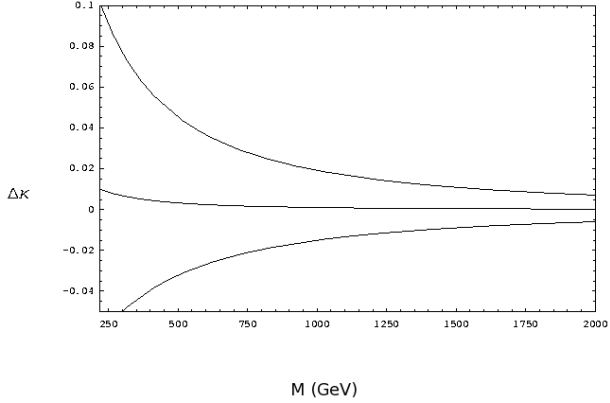}\\
  \caption{CMDM as  function of the topcolor particle masses: the curve at the top corresponds to $h^0_t$, the one at the middle to $\pi^+_t$ and the bellow to $\pi_t^0$}\label{tc2tres}
\end{figure}

\begin{figure}
  \includegraphics[width=10cm]{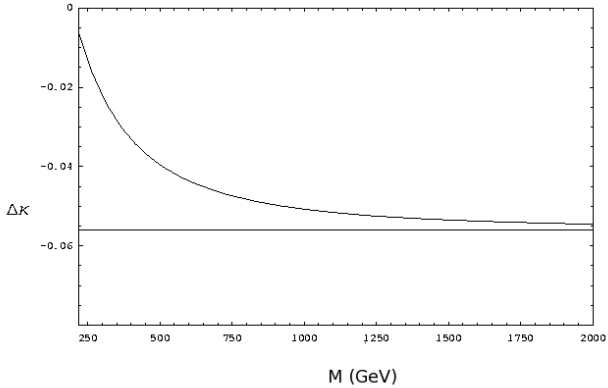}\\
  \caption{CMDM as function of the topcolor  particle mass: the horizontal line corresponds to the SM prediction and the line at the top gives the total topcolor TC2 predictions.}\label{tc2tresSM}
\end{figure}

In Figure \ref{tc2tres} we present the evolution of each one of
the topcolor scalars $h^0_t, \pi^\pm_t, \pi^0_t$ contributions to
the CMDM as function of their masses. In Figure \ref{tc2tresSM}
we compare the topcolor and the SM contributions to the top-quark
CMDM.

For $m_{h^0_t} \approx m_{\pi_t} \approx m_{\pi_0}\approx 250$ GeV
and summing the respective value for the SM, we find
\begin{equation}
\Delta\kappa^{TC2}_{(s)}=-0.01,
\end{equation}
which is of the order of the sensibility of the LHC.

There is a previous estimate of the contribution to the CMDM
induced by a techniscalar in TC2 models \cite{tres}. However, this
estimate did not consider the actual suppression factors
$(4\pi)^{-2}$ involved in one-loop calculations and it was
obtained a rather large value for this contribution, of order
0.1, even for a relatively massive techniscalar (0.5 TeV) \cite{tres}.

\subsection{Universal Extra Dimensions}

We consider a generalization of the SM where all the particles
propagate in five dimensions: $x = 0, 1, 2, 3$ correspond to the
usual coordinates and $x^5 = y$ is the fifth one. This extra
dimension will be compactified in a circle of ratio $R$ with the
points $y$ and $-y$ identified in a $S^1/Z_2$ orbifold
\cite{veintiuno,veintidos}. The terms that will contribute at
the one-loop order to the CMDM are given by
\begin{equation}  L = \int d^4x dy ({\cal L}_F
+ {\cal L}_Y) \label{lag5D}
\end{equation}
with
\begin{eqnarray}
{\cal L}_F &=& \left[\overline{Q} i \Gamma^M D_MQ + \overline{U} i
\Gamma^M D_M U +
\overline{D} i \Gamma^M D_M D\right],\nonumber\\
{\cal L}_Y &=& - \overline{Q} \lambda_u \widetilde{\Phi} U -
\,\overline{Q} \lambda_d \Phi D + h.c. \label{lagrangiano}
 \end{eqnarray}

The numbers $M, N = 0, 1, 2, 3, 5$ denote the five dimensional
Lorentz indexes. The covariant derivative is defined as $D_M
\equiv
\partial_M + i \widehat{g} W^a_M T^a + i \widehat{g}\, ' B_M Y$. The five dimensional gamma matrices $\Gamma_M$
 are $\Gamma_\mu = \gamma_\mu$ and $\Gamma_4 = i \gamma_5$ with the
 metric tensor given by $g_{M N} = ( +, -, -, -, - )$.

The Fourier expansions of the fields are given by:
\begin{eqnarray} B_\mu (x,y) &=& \frac{1}{\sqrt{\pi R}} B_\mu^{(0)}(x) +
\frac{\sqrt{2}}{\sqrt{\pi R}}\sum_{n=1}^\infty B_\mu^{(n)}(x)
\cos\left(\frac{n\pi y }{R}\right), \nonumber\\ B_5 (x,y) &=&
\frac{\sqrt{2}}{\sqrt{\pi R}}\sum_{n=1}^\infty B_5^{(n)}(x)
\sin\left(\frac{n\pi y }{R}\right),
\\
 Q(x,y) &=& \frac{1}{\sqrt{\pi R}} Q_L^{(0)}(x) +
\frac{\sqrt{2}}{\sqrt{\pi R}}\sum_{n=1}^\infty \left[Q_L^{(n)}(x)
\cos\left(\frac{n\pi y }{R}\right) + Q_R^{(n)} \sin
\left(\frac{n\pi y }{R}\right)\right], \nonumber\\ U(x,y) &=&
\frac{1}{\sqrt{\pi R}} U_R^{(0)}(x) + \frac{\sqrt{2}}{\sqrt{\pi
R}}\sum_{n=1}^\infty \left[ U_{t\; R}^{(n)}(x) \cos\ \left(
\frac{n\pi y }{R}\right) + U_{t\; L}^{(n)} \sin \left(\frac{n\pi
y }{R}\right)\right] \nonumber
\end{eqnarray}

The expansions for $B_\mu$ and $B_5$ are similar to the expansions
for all the gauge fields and the Higgs doublet (but this last one
without the $\mu$ or $5$ Lorentz index). It is by integrating the
fifth $y$ component that we obtain the usual interaction terms
and the KK spectrum for ED models, $Q^{(n)}=\left(Q^{(n)}_t \;\;\;
Q^{(n)}_b \right)^T$.

\begin{table}
\begin{center}
\begin{tabular}{||l||c|c|c|c||}
  \hline\hline
              & $A$ & $B$ & $A^{'}$  & $B^{'}$  \\
  \hline\hline
  $t^{(0)}A^{(n)}_\mu Q_t^{(n)}$ & $i\frac{2}{3} e$ & 0 & $i\frac{2}{3} e$ & 0 \\
  \hline
  $t^{(0)}Z^{(n)}_\mu Q_t^{(n)}$ & $i\frac{g}{c_W}(\frac{1}{2}-\frac{2}{3}s_W^2)$  & 0 & $i\frac{g}{c_W}(\frac{1}{2}-\frac{2}{3}s_W^2)$  & 0 \\
  \hline
  $t^{(0)}W^{(n)}_\mu Q_b^{(n)}$ & $i\frac{g}{\sqrt{2}}$ & 0 & $i\frac{g}{\sqrt{2}}$ & 0 \\
  \hline
  $t^{(0)}A^{(n)}_\mu U_t^{(n)}$ & 0 & $i\frac{2}{3} e$ & 0 & $i\frac{2}{3} e$ \\
  \hline
  $t^{(0)}Z^{(n)}_\mu U_t^{(n)}$ & 0 &  $i\frac{2}{3} g \frac{s_W^2}{c_W}$ & 0 &  $i\frac{2}{3} g \frac{s_W^2}{c_W}$ \\
  \hline\hline
\end{tabular}
\end{center}
\caption{Feynman rules for gauge KK of the UED contributions to the
CMDM of top quark}\label{UEDgauge}\end{table}

We will be interested in the third family of quarks and
$Q_t^{(n)}$ and $Q_b^{(n)}$ will refer to the upper and lower
parts of the doublet $Q$. Similarly, the $U_t^{(n)}$ and
$D_t^{(n)}$ will be the KK modes of the usual right-handed
singlet top quarks. There is a mixing between the masses of the
gauge eigenstates of the KK top quarks ($Q_t^{(n)}$ and
$U_t^{(n)}$), where the mixing angle is given by $
 \tan (2 \alpha_t^n) =
m_t/m_n \label{mix}$ with $m_n \approx n/R$. However, we will
neglect this mixing angle $(\sin \alpha_t = 0)$ for the purpose of
the present calculation.

 \begin{table}
\begin{center}
\begin{tabular}{||l||c|c|c|c||}
  \hline\hline
    & $A$ & $B$ & $A^{'}$ & $B^{'}$ \\
  \hline\hline
  $t^{(0)}A^{(n)}_5Q_t^{(n)}$ & $-\frac{2}{3} e$ & 0 & 0 & $\frac{2}{3} e$  \\
  \hline
  $t^{(0)}Z^{(n)}_5Q_t^{(n)}$ & $-\frac{g}{c_W}(\frac{1}{2}-\frac{2}{3}s_W^2)$ & 0  & 0 & $\frac{g}{c_W}(\frac{1}{2}-\frac{2}{3}s_W^2)$ \\
  \hline
  $t^{(0)}W^{(n)}_5Q_b^{(n)}$ & $-\frac{g}{\sqrt{2}}$ & 0 & 0 & $\frac{g}{\sqrt{2}}$  \\
  \hline
  $t^{(0)}A^{(n)}_5U_t^{(n)}$ & 0 & $\frac{2}{3}e$ & $-\frac{2}{3}e$  & 0\\
  \hline
  $t^{(0)}Z^{(n)}_5U_t^{(n)}$ &  $-\frac{2}{3} g \frac{s_W^2}{c_W}$ & 0 & 0 & $\frac{2}{3} g \frac{s_W^2}{c_W}$ \\
  \hline\hline
\end{tabular}
\end{center}
\caption{Feynman rules for KK scalar of the UED contributions to
the CMDM of top quark}\label{UEDscalar}\end{table}

\begin{table}
\begin{center}
\begin{tabular}{||l||c|c|c|c||}
  \hline\hline
              & $A$ & $B$ & $A^{'}$  & $B^{'}$  \\
  \hline\hline
  $t^{(0)}A^{a,(n)}_\mu Q_t^{(n)}$ & $i g_s$ & 0 & $i g_s$ & 0 \\
  \hline
  $t^{(0)}A^{a,(n)}_\mu U_t^{(n)}$ & 0 & $i g_s$ & 0 & $i g_s$  \\
  \hline
  $t^{(0)}A^{a,(n)}_5 Q_t^{(n)}$ & $g_s $ & 0 & 0 & $-g_s$ \\
  \hline
  $t^{(0)}A^{a,(n)}_5 U_t^{(n)}$ & 0 & $ -g_s$ &  $ g_s$ & 0 \\
  \hline\hline
\end{tabular}
\end{center}
\caption{Feynman rules for the KK states for the QCD contributions
to the CMDMT}\label{UEDgluon}\end{table}

In table V we include the Feynman rules for the fifth component of a
gauge field, the KK scalar which will induce the UED contribution to
the CMDM of the top quark (Figure 1a). Table VI contains the
respective Feynman rules for the couplings of the external quark $t$
which corresponds to the $t^{(0)}$ component with a gauge and 
quark excitations (Fig. 1a). Table VII includes the respective Yukawa
couplings for an external quark $t^{(0)}$ with the scalar $KK$
states (Fig. 1a). Finally, the couplings presented in Table 8 will
be used to compute the QCD contribution to the CMDM induced by the
$KK$ excitations associated to the fifth component of the gluon
field $A^{Q,(n)}_5$ (Figs. 1a and 2a).

Here we will take $1/R \approx 1$ TeV. The $KK$ modes corresponding
to the EW sector for the gauge (Table (\ref{UEDgauge})) and scalar
(Table (\ref{UEDscalar})) in the loop, respectively, then give the
following contribution to the CMDM
\begin{eqnarray}
\Delta\kappa_{(\mu)}^{EW,5D}&=&-1.3\times 10^{-4}  \nonumber \\
\Delta\kappa_{(s)}^{EW,5D}&=& -4.1 \times 10^{-4}
\end{eqnarray}
with a total value for the electroweak contribution given by
\begin{equation}
\Delta\kappa^{EW,5D}= -5.4\times 10^{-4} \label{EW5D}
\end{equation}
The KK modes of the QCD sector induces the following contributions
\begin{eqnarray}
\Delta\kappa_{(\mu)}^{QCD,5D}&=& -9.4\times 10^{-4} \nonumber \\
\Delta\kappa_{(s)}^{QCD,5D}&=& -8.2\times 10^{-4}
\end{eqnarray}
corresponding to the gauge and scalar excitations contribution,
respectively, which add to the following value
\begin{equation}
\Delta\kappa^{QCD,5D}= -1.8\times 10^{-3} \label{QCD5D}
\end{equation}
Adding the EW eq. (\ref{EW5D}) and QCD eq. (\ref{QCD5D})
contributions we get finally
\begin{equation}\Delta\kappa^{5D}= -2.3\times 10^{-3}.\end{equation}

\section{CONCLUDING REMARKS}

In conclusion, we have made a critical analysis of the one loop
calculations for the CMDM of the top quark model. 
Our new results differ slighty from those already published for the
SM  and the THDM \cite{nueve}. In the TC2 model, our one-loop computation
shows a smaller value for the CMDM then the one previously obtained
\cite{tres}, which in turn is also in agreement with the constraints obtained
from low-energy precision experiments \cite{nueve}. As far as we know, the
one-loop calculations for the 331 and universal extra-dimensions
models have not been performed previously. In particular, the 331
result is suppressed by two orders of magnitud with respect to the
SM result and the extra-dimensions result is also in agreement
with the low-energy precision contraints. In this respect, a
precise measurement of the top-quark CMDM may be used to
distinguish between competing extensions of the SM. If we sum to
the SM CMDM the additional contributions given for the different
models considered in the present paper, we get the following
results
\begin{eqnarray}
\Delta\kappa^{SM}&=&-5.6\times 10^{-2} \;\; , \;\;
\Delta\kappa^{THDM}=-5.1\times 10^{-2} \nonumber \\
\Delta\kappa^{TC2}&=&-1.0\times 10^{-2} \;\; , \;\;
\Delta\kappa^{5D}=-5.8\times 10^{-2}.
\end{eqnarray}

Finally, eventhough the effects induced at NLO QCD corrections on
top quark production and decays have already been studied
\cite{veintitres}, as far as we know the complete NLO QCD
corrections to the CMDM $\Delta\kappa$ of the top quark have not
been calculated in any of these models.  It is important also to point 
out that the angular distributions of leptons or jets due to $t\bar t$ spin 
correlations allow a determination of the CMDM of the top quark with an 
accuracy of order 0.1 \cite{veintecinco}. This method provides a competitive 
way to observe new physics contributions to the CMDM, which is stable 
against experimental uncertainties \cite{veintecinco}.

\newpage

\appendix

\section{CMDM ANAYLTICAL EXPRESSIONS}

 In this appendix we present the analytical expressions obtained
    for each one-loop Feynman diagram involved in the calculation
    of the top-quark CMDM. We have used these expressions in order
    to perform the respective numerical calculations that lead to
    the CMDM results presented in this paper. This method has already
    been used to compute higher order corrections to fermion vertices
    or flavor-changing neutral vertices involving  leptons or
    the top quark \cite{Deshapande}.

The Feynman diagram shown on Fig. 1a corresponds to the scalar
contribution to the top-quark CMDM. In the SM, the scalars
circulating in the loop can be $h^0$, $G_Z$, and $G_W$, while in the
331 models it is $G_X$. In the UED models, this contribution arises
from the scalar $KK$ excitations $h^{0,(n)}, G^{(n)}_Z, ....$, and
fifth component of the gauge fields $A^{(n)}_5, Z^{(n)}_5,
W^{(n)}_5$. The respective QCD excitations for the $KK$ sector
correspond to $A^{a,(n)}_5$. We will use the following notation for
the incoming and the outcoming scalar couplings involved in Fig. 1a,
respectively
\begin{equation} (A P_L + B P_R) T^a \;\;\; ,\;\;\; (A'
P_L + B' P_R) T^a
\end{equation}
where $T^a$ corresponds to the $SU(3)_C$ generators in the case of
QCD scalar loop and it is just the unit matrix otherwise.

The contribution to the CMDM arising from the diagram given in Fig.
1a is thus given by the expression,
\begin{eqnarray}
\Delta\kappa^{QCD}_{(s)}&=&-\frac{1}{6\times 8 \pi^2}\int_0^1
dx\int_0^{1-x}dy
\frac{1}{(x+y)^2+(x+y)(\widehat{M}^2-\widehat{m}^2-1)+\widehat{m}^2}\nonumber
\\
&\times& \{(x+y)(x+y-1)(B'A+A'B)-\widehat{M}(x+y)(A'A+B'B)\}
\end{eqnarray}
where $m (M)$ correspond to the scalar (fermion) mass in the loop
and the factor (-1/6) comes from the $SU(3)_C$ generator algebra,
and $\widehat {m}=m/m_t$ is the normalized top-quark mass. In order
to get the respective EW scalar contribution for Fig. 1a, we have to
replace the factor (-1/6) by the unit and take the appropriated
scalar couplings.

In particular, if we set $ A = B = A^{'} = B^{'} = 1 $ and
$\widehat{M}=1$ in Eq. (A.2), we get directly the  $h^0$
contribution to  $\Delta\kappa_{(5)}$ which agrees with the
result given in Eq. (A.10) of Fujikawa et al.
\cite{Fujikawa}. On the other hand, if we set $ A = A^{'} = -
B = - B^{'} = 1 $ and $\widehat{M}=1$  in our Eq. (A.2), we get
the respective pseudoscalar $G_Z$ contribution which also agrees
with Eq. (3.5) of Ref. \cite{Fujikawa}.

If we have two $KK$ excitations running in the loop of Fig. 1a,
the respective analytical expression is given by
\begin{eqnarray}
\Delta\kappa_{(s)}&\approx & -\sum_{n=1}^{\infty}\frac{1}{12\times
48 \pi^2}\frac{1}{\widehat{(n/R)}^2} \{3(B'A+A'B)-2
\widehat{M}(A'A+B'B)\}
\nonumber \\
&\approx& -\frac{\widehat{R}^2}{72\times 48} \{3(B'A+A'B)-2
\widehat{M}(A'A+B'B)\},
\end{eqnarray}
where we have used the approximation that the masses of both
excitations are of the same order of magnitude $m_n\sim n/R$ and
we have neglected any other mass circulating in the loop.

The diagram shown in Fig. 1b receives contributions from the SM
gauge bosons W, Z, g, the 331 gauge boson X, and for the UED model
it can be the respective EW or QCD gauge bosons $A^{(n)}_\mu,
Z^{(n)}_\mu, W^{(n)}_\mu$ and $A^{a,(n)}_\mu$. In this case, we use
the following notation for the gauge and fermionic couplings
involved in this diagram,
\begin{equation}
\gamma_\alpha (A P_L + B P_R)T^a \;\;\;\;\;\; \gamma_\alpha (A'
P_L + B' P_R)T^a,
\end{equation}
where the $T^a$ correspond again to the $SU(3)_C$ generators. The
respective analytical expression for this contribution is given
by,
\begin{eqnarray}
\Delta\kappa_{(\mu)}&=&-\frac{1}{6\times 4 \pi^2}\int_0^1
dx\int_0^{1-x}dy
\frac{1}{(x+y)^2+(x+y)(\widehat{M}^2-\widehat{m}^2-1)+\widehat{m}^2}\nonumber
\\
& \times&\{(x+y-2)(x+y-1)(A'A+B'B)\nonumber \\
&-&\widehat{M}(x+y-1)(A'B+B'A)\}.
\end{eqnarray}

\begin{eqnarray}
\Delta\kappa_{(\mu)}&\approx&
-\sum_{n=1}^{\infty}\frac{1}{12\times 24
\pi^2}\frac{1}{(\widehat{n/R})^2} \{3(A'A+B'B)-2
\widehat{M}(A'B+B'A)\} \nonumber \\
&\approx& -\frac{\widehat{R}^2}{72\times 24} \{3(A'A+B'B)-2
\widehat{M}(A'B+B'A)\},
\end{eqnarray}
In order to get the EW contribution for the diagram 1b, we have to
make the same sustitutions indicated for the expression (A.2).

For example, if we apply Eq. (A.5) to the loop induced
by the exchange of a $Z$ gauge boson and set $\widehat{M}=1$, we
get direclty Eq. (A.4) of Fujikawa et al. \cite{Fujikawa}
with $\eta=1$.  For the loop induced by the exchange of a gluon,
if we set $\widehat{M}=1$ $\widehat{m}=0$, and $A = B = A^{'} =
B^{'} = -i g_s$, we obtain
\begin{equation}
\Delta\kappa^{QCD}_{(\mu)}=-\frac{1}{6}\frac{\alpha_s}{\pi}.
\end{equation}
The above result reduces to the well known QED result
$\Delta\kappa=\alpha/(2\pi)$ if we supress the factor $-1/6$
coming from $SU(3)_C$ algebra and an additional $1/2$ factor
which was introduced in the definition of $\Delta\kappa$ given in
Eq. (1).

In the UED models, the diagram shown in Fig. 2a receives
contributions from a colored scalar particle which corresponds to
the dimension 5 component of the gluons, $A^{a,(n)}_5$. In this
case we use the following notation for the respective couplings,
\begin{equation}
(A P_L + B P_R)T^c \;\;\; , \;\;\;  (A' P_L + B' P_R) T^d.
\end{equation}
where the couplings for the excitated colored scalar with the
external gluon may be taken from Ref.\cite{diecinueve}. The
corresponding contribution has the analytical expression,
\begin{eqnarray}
\Delta\kappa_{(s)}&=&\frac{3}{16\pi^2}\int_0^1
dx\int_0^{1-x}dy
\frac{1}{(x+y)^2+(x+y)(\widehat{M}^2-\widehat{m}^2-1)+\widehat{m}^2}
\nonumber\\
& \times &\{(x+y)(x+y-1)(A'B+B'A)\nonumber \\
&-&\widehat{m}(x+y-1)(A'A+B'B)\}
\end{eqnarray}

and in this case the factor 3/2 comes from the Lie algebra of the
generators $f^{abc}T^bT^c$. The expression for the $KK$
contributions in this case reduces to

\begin{eqnarray}
\Delta\kappa_{(s)}&\approx&
-\sum_{n=1}^{\infty}\frac{3}{12\times 16
\pi^2}\frac{1}{(\widehat{n/R})^2} \{3(A'B+B'A)+2
\widehat{M}(A'A+A'A)\} \nonumber \\
&\approx& \frac{\widehat{R}^2}{24\times 16} \{3(A'B+B'A)+2
\widehat{M}(A'A+B'B)\},
\end{eqnarray}
where we have assumed again that the the excitation masses
running in the loop are of the same order of magnitude.

The diagram given in Fig. 2b receives contributions from the gluons
$A^a_\mu$ or the respective $KK$ excitations $A^{a,(n)}_\mu$, the
notation for the respective couplings is given by

\begin{equation}
\gamma_\alpha (A P_L + B P_R)T^a \;\;\;\;\;\; \gamma_\beta (A'
P_L + B' P_R)T^b
\end{equation}
and thus the corresponding QCD contribution to the CMDM may be
expressed as

\begin{eqnarray}
\Delta\kappa_{(\mu)}&=&\frac{3}{8 \pi^2}\int_0^1
dx\int_0^{1-x}dy
\frac{1}{(x+y)^2+(x+y)(\widehat{M}^2-\widehat{m}^2-1)+\widehat{m}^2}\nonumber
\\
& \times&\{(x+y)(x+y-1)(A'A+B'B)\nonumber \\
&-&\widehat{M}(x+y-1)(A'B+B'A)\}
\end{eqnarray}

Finally, if we have two $KK$ excitations running in this loop, we
get the following expression for this contribution

\begin{eqnarray}
\Delta\kappa_{(\mu)}&=&\frac{3}{8 \pi^2}\int_0^1
dx\int_0^{1-x}dy
\frac{1}{(x+y)^2+(x+y)(\widehat{M}^2-\widehat{m}^2-1)+\widehat{m}^2}\nonumber
\\
& \times&\{(x+y)(x+y-1)(A'A+B'B)\nonumber \\
&-&\widehat{M}(x+y-1)(A'B+B'A)\}.
\end{eqnarray}

\noindent {\bf Acknowledgments} \\

This work was supported by Fundaci\'on Banco de la
Rep\'ublica (Colombia), CONACyT (M\'exico) and the HELEN proyect. \\

\end{document}